\documentclass[%
reprint,
amsmath,amssymb,
aps,
]{revtex4-1}
\usepackage{xcolor}
\usepackage{soul}
\usepackage{textcomp}
\usepackage{graphicx}
\usepackage{dcolumn}
\usepackage{bm}

\begin{document}
	
	\preprint{APS/123-QED}
	
	\title{Fluctuations and correlations of transmission eigenchannels in diffusive media}

	\author{Nicholas Bender$^{1}$}
	\thanks{These two authors contributed equally.}
	
	\author{Alexey Yamilov$^{2},^{*}$}
    \email{yamilov@mst.edu}

	\author{Hasan Y\i lmaz$^{1}$}
	
	\author{Hui Cao$^{1}$}
	\email{hui.cao@yale.edu}
	
	\affiliation{Department of Applied Physics, Yale University, New Haven, Connecticut 06520, USA \\
		Physics Department, Missouri University of Science \& Technology, Rolla, Missouri 65409, USA}

	\date{\today}
	
	\begin{abstract}
		
    Selective excitation of a diffusive system’s transmission eigenchannels enables manipulation of its internal energy distribution. The fluctuations and correlations of the eigenchannels’ spatial profiles, however, remain unexplored so far. Here we show that the depth profiles of high-transmission eigenchannels exhibit low realization-to-realization fluctuations. Furthermore, our experimental and numerical studies reveal the existence of inter-channel correlations, which are significant for low-transmission eigenchannels. Because high-transmission eigenchannels are robust and independent from other eigenchannels, they can reliably deliver energy deep inside turbid media.	
		
	\end{abstract}
	
	\maketitle
	
	
	In recent years, extensive studies of coherent wave transport in multiple-scattering media have been conducted with light, microwaves, and acoustic waves~\cite{2012_Mosk_SLM_review, 2017_Rotter_Gigan_review}. The overarching goal of this research is overcoming the limitations imposed by incoherent diffusion: thereby enabling energy delivery deep inside a turbid medium. While multiple scattering persistently randomizes waves traveling in a linear system with static disorder, the coherent wave transport is ultimately a deterministic process. Therefore, it can be described by a field transmission matrix $t$, which maps the incident waves to the transmitted waves~\cite{1997_Beenakker}. The eigenvectors of $t^{\dagger}t$ provide the input wavefronts which excite a set of disorder-specific wavefunctions –spanning the system– known as the transmission eigenchannels. Any incoming wave can be decomposed into a linear combination of eigenchannels, each propagating independently through the system with a transmittance given by the corresponding eigenvalue $\tau$. 		One of the striking theoretical predictions of diffusive systems is the bimodal distribution of the transmission eigenvalues: with maxima at $\tau=0$ and $\tau=1$~\cite{1984_Dorokhov, 1986_Imry, 1988_Mello, 1990_Pendry_MacKinnon, 1996_Nazarov_Eigenvalues}. The corresponding eigenchannels are referred to as closed and open channels.

	Both the fluctuations of, and the correlations between transmission eigenvalues are intensely studied topics ~\cite{1997_Beenakker, 2017_Rotter_Gigan_review, 1990_Pichard, shi2012transmission}. This fundamental research area has provided explanations for prominent physical phenomena like universal conductance fluctuations and quantum shot noise ~\cite{1986_Imry, 1988_Feng, 1991_Altshuler, 1993_Beenakker, 1994_Berkovits_Feng, 1995_Caselle, 1996_Nazarov_Eigenvalues, 1997_Beenakker, 2011_Beenakker}. However, the statistical properties of individual eigenchannels, such as the fluctuations of eigenchannel profiles and correlations between them, have not been studied before. In electronic systems, this is because input states cannot be easily controlled and therefore systematically exciting individual eigenchannels is unfeasible. Thanks to the recent developments of optical wavefront shaping techniques, photonic systems offer a unique opportunity for studying the second-order statistics of transmission eigenchannels.
	
    The ability to manipulate input states in optics and acoustics has spurred a renewed interest in using transmission eigenchannels for imaging and sensing applications~\cite{2012_Mosk_SLM_review,2015_Vellekoop_review,2015_Yu_Wavefront_Shaping_Review,2017_Rotter_Gigan_review}. Coupling waves into an open channel, not only enhances the transmitted power through a diffusive system~\cite{2008_Vellekoop_PRL, 2012_Kim_NatPho, 2013_Yu_PRL, 2013_Kim_OL, 2014_Popoff_PRL, 2016_Bosch_OE, 2017_Hsu_PRL}, but also enhances the energy density inside the system~\cite{2011_Mosk, 2014_Gerardin_Full_Matrix, 2014_Chabanov_NatComm, 2015_Genack_Eigenchannels_Inside, sarma2015control, 2016_Ojambati_Fundamental_Mode_Experiment, SarmaPRL16, 2018_Hong_Optica, 2019_Yilmaz_Transverse_localization}. The latter has a tremendous impact on enhancing light-matter interactions and manipulating nonlinear processes in turbid media~\cite{2016_Liew_ACS_Photon, 2018_Bertolotti_Raman}. So far, however, the potential energy density enhancement is only known after ensemble averaging over many disorder realizations. Thus, it is still an open question if coupling energy into an open channel guarantees a significant enhancement of the energy density inside a single diffusive sample.
		
	Here, we experimentally and numerically investigate both the fluctuations and correlations of transmission eigenchannel depth profiles in optical diffusive systems. We develop novel experimental techniques for measuring the transmission matrix of an on-chip diffusive waveguide, exciting its individual transmission eigenchannels, and performing an interferometric measurement of the light field everywhere inside the waveguide. High-transmission eigenchannels exhibit small realization-to-realization fluctuations in their depth profiles, demonstrating a robustness when compared to either low-transmission eigenchannels or random inputs. Furthermore, different eigenchannels are correlated in their depth profile fluctuations from realization-to-realization. The correlations are weaker for higher-transmission eigenchannels, indicating they are more independent than lower-transmission eigenchannels. Their consistent depth profiles guarantee deep penetration of energy into any diffusive system, which is promising for applications in deep tissue imaging and light delivery.

	\begin{figure}
		\begin{center}
			\includegraphics[width=\linewidth]{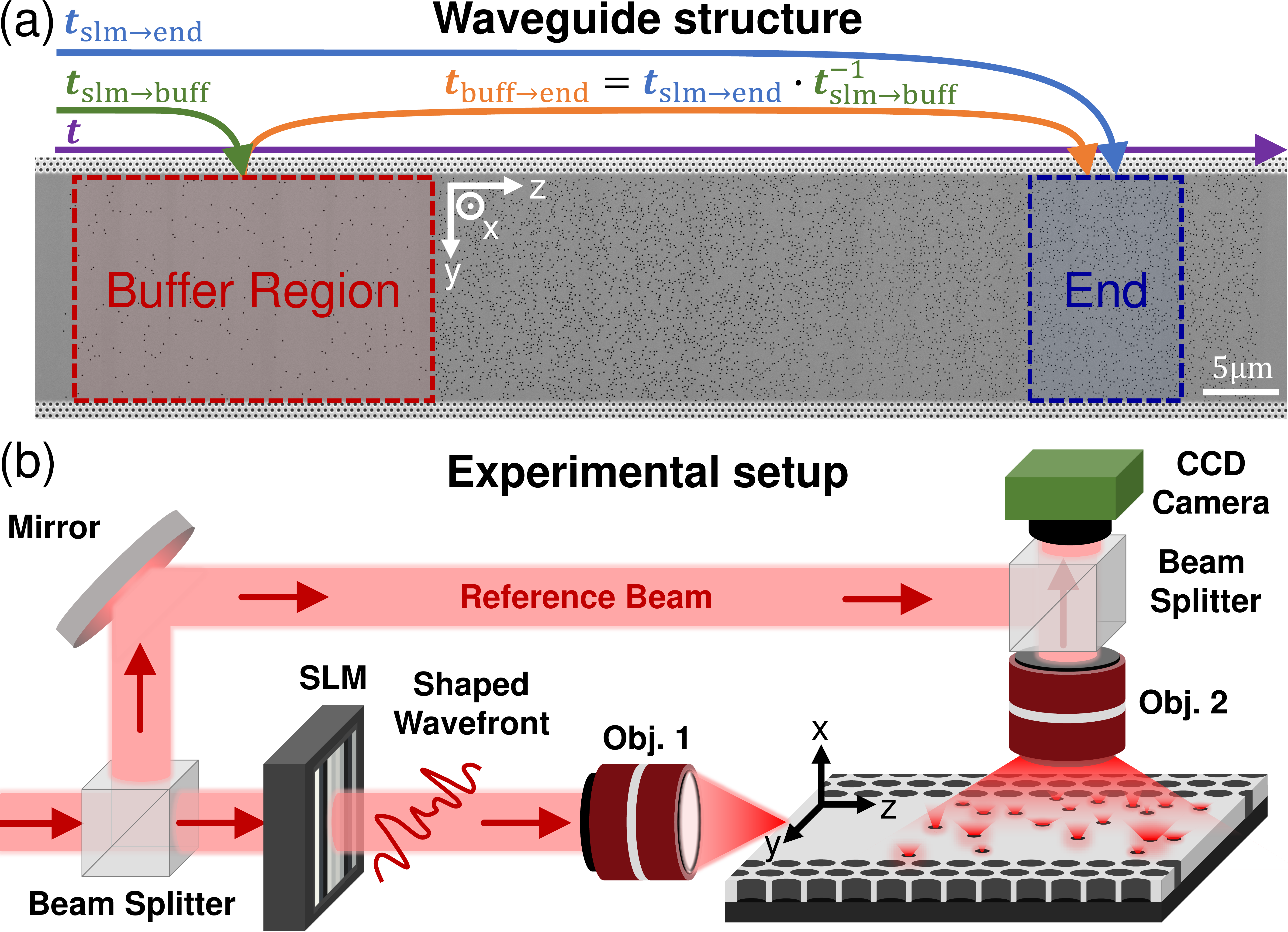}
			\caption{\label{fig1} 
				Waveguide structure and optical setup. A composite SEM image of a diffusive waveguide is shown in (a). The matrix mapping the field in the buffer region to the end region, $t_{\rm buff\rightarrow end}$, is related to the matrices $t_{\rm slm \rightarrow \rm buff}$ and $t_{\rm slm \rightarrow \rm end}$. In (b) the simplified sketch of the experimental setup illustrates how we wavefront shape a laser beam with a spatial light modulator (SLM) while performing an interferometric measurement of the light scattered out of the waveguide.
			} 
		\end{center}
	\end{figure}

	To directly observe the depth profiles of transmission eigenchannels \textit{within} a diffusive system, we fabricate two-dimensional (2D) waveguide structures on a silicon-on-insulator wafer with electron beam lithography and plasma etching~\cite{SM}. As shown in Fig.~\ref{fig1}(a), 100 nm-diameter holes are randomly etched into the waveguides, which have photonic crystal sidewalls to reflect light~\cite{2014_Yamilov_Dofz_experiment}. At the wavelength of our probe light, $\lambda = 1.55$ \textmu m, the transport mean free path, $\ell_t = 3.2$ \textmu m, is much shorter than the disordered region length, $L = 50$ \textmu m, in each waveguide. Therefore, the light undergoes multiple scattering and diffusive transport through each waveguide~\cite{SM}. Light scatters out-of-plane from the random holes, providing a direct probe of the light inside the disordered region. This process can be modeled as an effective loss, and accounted for in the diffusive dissipation length: $\xi_a$ = 28 \textmu m. The waveguides are each 15 \textmu m wide, supporting $N$ = 55 propagating modes at $\lambda = 1.55$ \textmu m. Before entering one of the diffusive waveguides, light is injected via the edge of the wafer into a ridge waveguide. Due to the large refractive index mismatch between silicon and air, only low-order waveguide modes are excited at the interface. Before the disordered region, the waveguide width is tapered from 300 \textmu m to 15 \textmu m in order to convert the lower-order modes to higher-order ones. The taper enables us to access all waveguide modes incident on the disordered region~\cite{SarmaPRL16}.  
	
	To measure the light field inside individual diffusive waveguides, we use an interferometric setup, as sketched in Fig.~\ref{fig1}(b). In our setup, the monochromatic light from a wavelength-tunable laser source is split into two beams. One beam is modulated by a spatial light modulator (SLM) and then injected into one of the waveguides via the edge of the wafer. The other beam is used as a reference beam. It is spatially overlapped with the out-of-plane scattered light from the diffusive waveguide: on the CCD camera chip. The CCD camera records the resulting interference pattern, from which the complex field profile across the diffusive waveguide is obtained, as shown in~\cite{SM}.
	
	By sequentially applying an orthogonal set of phase patterns to the 128 SLM macropixels, and measuring the field within the sample, we acquire a matrix that maps the field from the SLM to the field inside the disordered waveguide $t_{\rm slm \rightarrow \rm int}$. This matrix encompasses information about the light transport inside the waveguide and the light propagation from the SLM to the waveguide. To separate these, we need access to the field incident on the disordered region of the waveguide. We obtain this information by adding an auxiliary weakly-scattering region in front of the diffusive region called the `buffer' region: as shown in Fig.~\ref{fig1} (a). From the light scattered out-of-plane from the buffer, we recover the field right in front of the strongly-scattering region. The length of the buffer region is 25 \textmu m, which is shorter than its 32 \textmu m-length transport mean free path. Therefore, light only experiences single scattering in the buffer, and as a result, the diffusive wave transport in the original disordered region is not appreciably altered.
	
\begin{figure}[t]
		\centering
		\includegraphics[width=\linewidth]{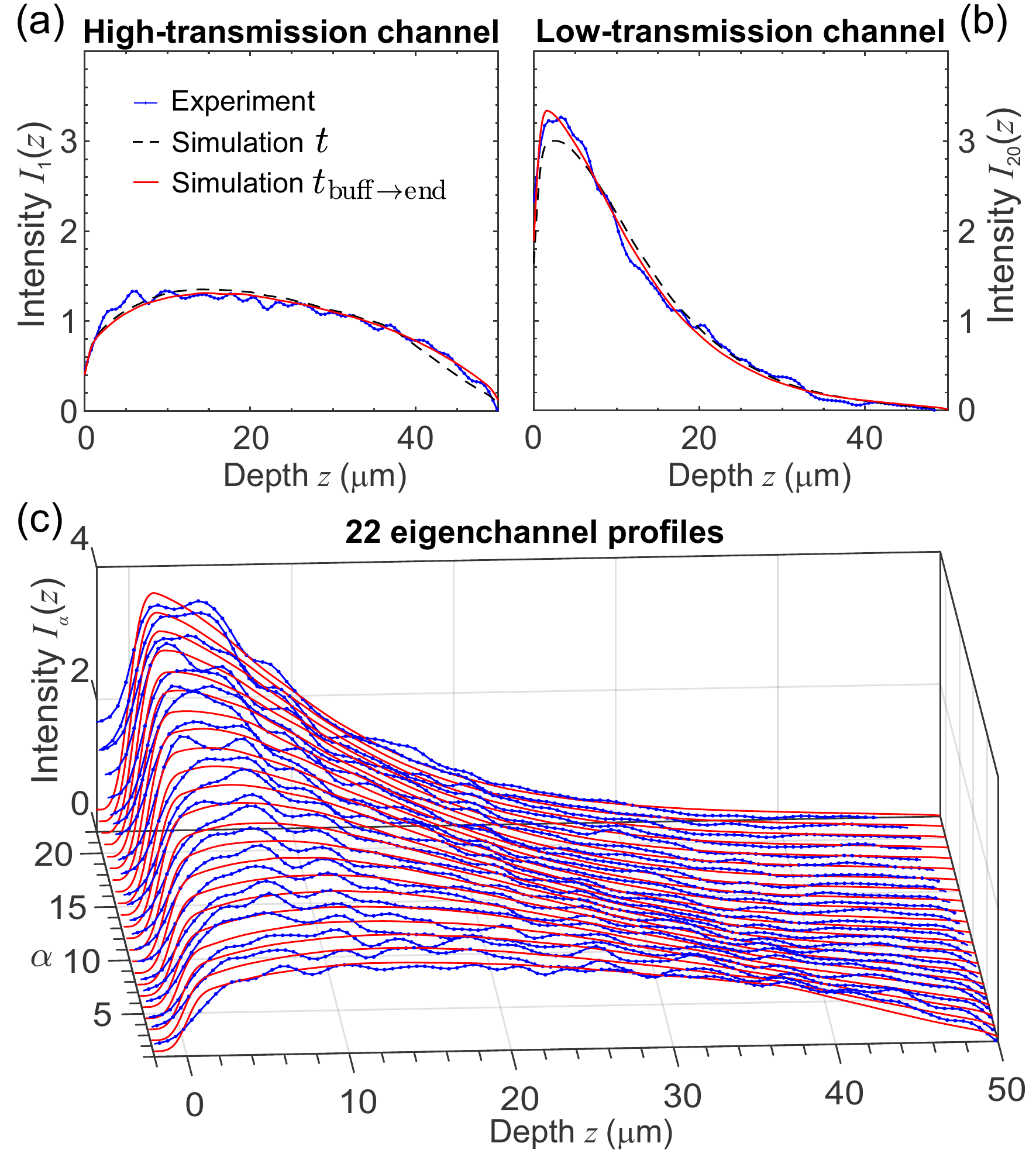}
		\caption{\label{profiles} 
			Depth profiles of transmission eigenchanels. High ($\alpha=1$) and low ($\alpha=20$) transmission eigenchannel profiles are presented in (a,b) while the 22 measured eigenchannel profiles are juxtaposed in (c). The experimentally measured profiles (blue lines) agree well with the profiles calculated from numerical simulations using the transmission matrix $t$ (black dashed lines) and the matrix $t_{\rm buff\rightarrow end}$ (red lines).
		} 
	\end{figure}

	With access to the field inside the buffer, we can construct the matrix relating the field on the SLM to the buffer, $t_{\rm slm \rightarrow \rm buff}$. From $t_{\rm slm \rightarrow \rm int}$, we can also construct the matrix, $t_{\rm slm \rightarrow \rm end}$, which maps the field from the SLM to a region near the end of the diffusive waveguide. With these we calculate the matrix which maps the field from the buffer to the end, $t_{\rm buff\rightarrow end}= t_{\rm slm\rightarrow end} \, t_{\rm slm\rightarrow buff}^{-1}$, using Moore-Penrose matrix inversion. Although $t_{\rm buff\rightarrow end}$ is not the field transmission matrix $t$, the depth profiles of its eigenchannels match those of transmission eigenchannels in our numerical simulation (see Fig.~\ref{profiles} and discussion below). Therefore, $t_{\rm buff\rightarrow end}$ can be used as an experimental proxy for the field transmission matrix, $t$, of the diffusive waveguide.
	
	To excite a single eigenchannel, we first perform a singular value decomposition on $t_{\rm buff\rightarrow end}$ to obtain the field distribution in the buffer corresponding to one eigenchannel. Then we multiply the field profile in the buffer with $t^{-1}_{\rm slm\rightarrow buff}$ to calculate the SLM phase-modulation pattern. By displaying this pattern on the SLM, we excite a single eigenchannel of the diffusive waveguide. We record the spatial intensity profile of each eigenchannel within the diffusive waveguide. From this measurement, we obtain the eigenchannel depth profile $\tilde{I}(z)$ associated with each measurement by summing the intensity over the waveguide cross-section. For each depth profile, the measured intensity profile $\tilde{I}(z)$ is normalized to $I(z)=\tilde{I}(z)/ [(1/L)\int_0^L \tilde{I}(z')dz']$.
	
	In Figures~\ref{profiles}(a) and (b), the experimentally-measured depth profiles of a high-transmission and a low-transmission eigenchannel are juxtaposed. The high-transmission eigenchannel in (a) has an arch-shaped energy-density distribution which spans the depth of the diffusive region. In (b), the energy-density distribution of the low-transmission eigenchannel rapidly decays with depth. We numerically calculate the transmission eigenchannels with the recursive Green’s function method in the Kwant simulation package \cite{SM}. The experimentally measured profiles match the corresponding depth profiles generated from numerical simulations of both $t$ and $t_{\rm buff\rightarrow end}$; confirming that we excite individual eigenchannels in our measurements. Furthermore, the agreement between the eigenchannels of $t_{\rm buff\rightarrow end}$ and $t$, confirms that the depth profiles of $t_{\rm buff\rightarrow end}$ have a one-to-one correspondence with the eigenchannels of $t$.  
	
	In total, we measure $50$ eigenchannel profiles for a single experimental system realization. Each profile matches one of the ensemble-averaged profiles of $t_{\rm buff\rightarrow end}$ generated numerically without any fitting parameters~\cite{SM}. Measurement noise causes multiple experimental profiles to be mapped to a single numerical profile, and this limits the total number of recovered eigenchannels to $22$. Fig.~\ref{profiles}(c) shows the depth profiles for all $22$ eigenchannels, which agree well with the numerical simulations~\cite{SM}. The transmittance of the measured eigenchannels varies from $\tau_1\simeq 0.43$ to $\tau_{22}\simeq 7.9\times 10^{-4}$, with a mean value of $\langle \tau_\alpha \rangle = 0.041$.

	\begin{figure}[htb]
		\centering
		\includegraphics[width=\linewidth]{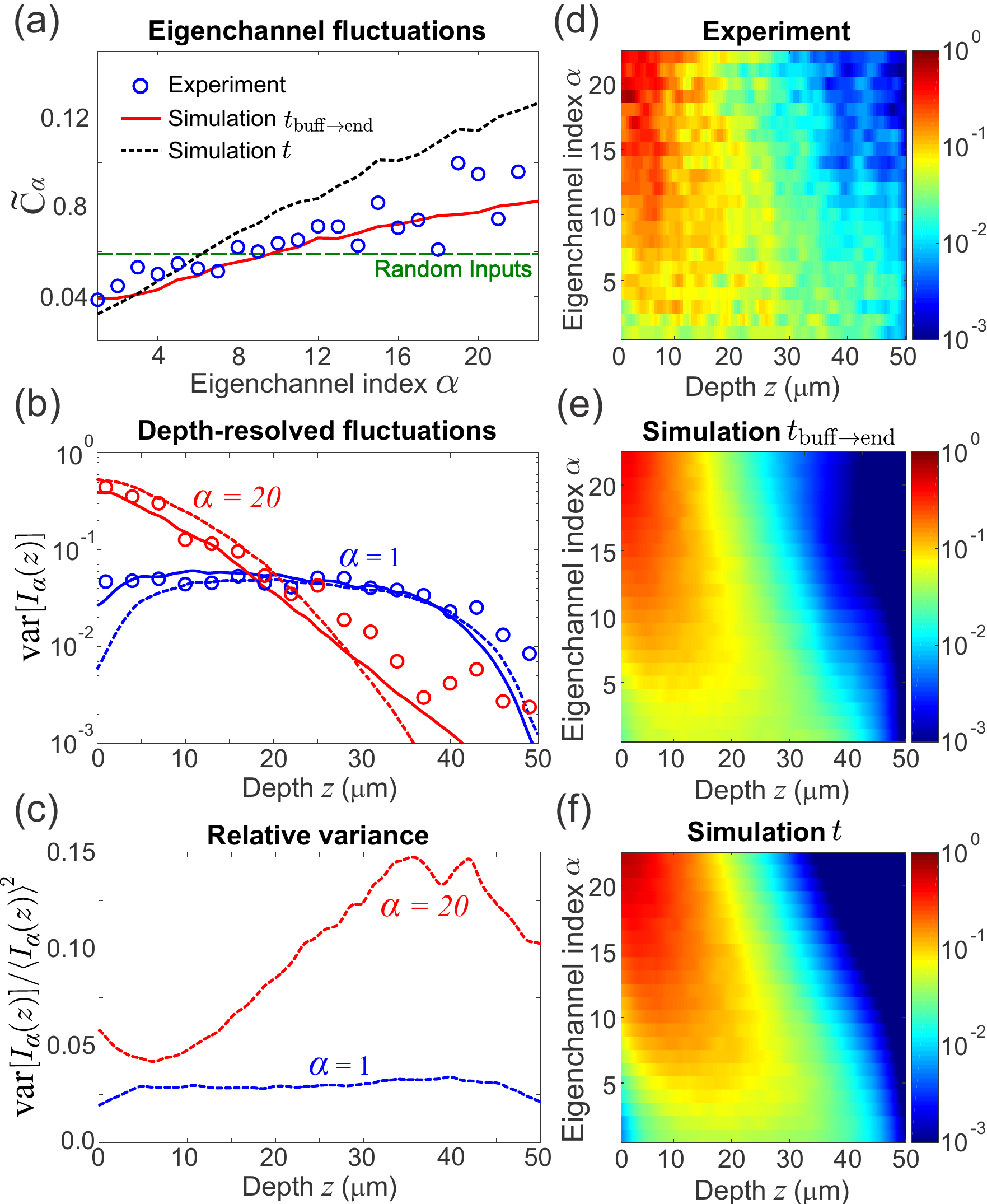}
		\caption{\label{fluctuations} 
			Eigenchannel fluctuations. In (a), the spatially-averaged depth-profile fluctuations of the eigenchannels, $\tilde{C}_\alpha$, increase monotonically with the channel index $\alpha$. The green dashed line indicates the experimentally observed fluctuations for random incident wavefronts: 0.59. In (b), experimentally observed depth-resolved intensity fluctuations, $ {\rm var}[I_\alpha (z)]$, of high ($\alpha=1$) and low ($\alpha=20$) transmission eigenchannels (circles) are closely reproduced by the numerical simulations of transmission eigenchannels from $t_{\rm buff\rightarrow end}$ (solid lines) and $t$ (dashed lines). In (c), $ {\rm var}[I_\alpha (z)]$ is divided by $\langle I_\alpha(z) \rangle^2$ for the high/low-transmission eigenchannels of $t$. In (d-e), the experimentally-observed and numerically-calculated depth-resolved intensity fluctuations for individual eigenchannels show how $ {\rm var}[I_\alpha (z)]$ evolves with $\alpha$.
		} 
	\end{figure}
	
	Next, we study the realization-to-realization fluctuations of eigenchannel profiles. From measurements of 13 system realizations~\cite{SM}, we compute the mean depth profile of each eigenchannel, $\langle I_\alpha (z)\rangle$, and the realization-specific deviation, $\delta I_\alpha (z) = I_\alpha (z) -\langle I_\alpha (z)\rangle$. From this, the total fluctuation of each eigenchannel profile is quantified by $\tilde{C}_\alpha = (1/L) \int_0^L \langle [\delta I_\alpha (z)]^2 \rangle dz$, where $\langle ...\rangle$ represents ensemble averaging. Fig.~\ref{fluctuations}(a) shows that the total fluctuation of each eigenchannel profile increases monotonically as a function of eigenchannel index. The uncertainty of $\tilde{C}_\alpha$ -due to the finite number of ensembles in our experiment- is estimated from simulations to be $\pm 25\%$ the value of $\tilde{C}_\alpha$: which is smaller than the overall change of $\tilde{C}_\alpha$ with $\alpha$. Hence, the depth profiles of high-transmission eigenchannels fluctuate less than the profiles generated by random illumination patterns (indicated by the green dashed line); while lower-transmission eigenchannels fluctuate more.  
	
    Now we look into the position-dependent fluctuation of individual eigenchannel profiles about their ensemble average, $ {\rm var}[I_\alpha (z)]= \langle [\delta I_\alpha (z)]^2 \rangle$ as a function of depth, $z$. Fig.~\ref{fluctuations}(b) reveals distinct differences in the depth dependence of high and low-transmission eigenchannels. While $ {\rm var}[I_\alpha (z)]$ is nearly flat for the high-transmission eigenchannel, it features a fast drop with $z$ for the low-transmission eigenchannel. Figs.~\ref{fluctuations}(d-f) are 2D plots of ${\rm var}[I_\alpha (z)]$ for all 22 eigenchannels: calculated using experimental data, as well as simulations of $t_{\rm buff\rightarrow end}$, and $t$. As the transmittance decreases, the maximum of $ {\rm var}[I_\alpha (z)]$ moves towards the front surface of the diffusive region. The decrease in the variance with depth results from the decay of the mean intensity with depth: $\langle I_\alpha (z)\rangle$. However, the relative intensity fluctuation of the low-transmission eigenchannels, characterized by ${\rm var}[I_\alpha (z)]/ \langle I_\alpha (z)\rangle^2$, actually increase with depth as shown in Fig.~\ref{fluctuations}(c) for $\alpha = 20$. In contrast, the relative intensity fluctuation of high-transmission eigenchannels is uniform with depth and small: for example ${\rm var}[I_1 (z)]/ \langle I_1 (z)\rangle^2 < 0.04$ for all $z$. Moreover, the fluctuation of a transmission eigenchannel's intensity at the sample output reflects the fluctuation of the corresponding transmission eigenvalue. Therefore, the stronger fluctuation of a low-transmission eigenchannel, relative to a high-transmission eigenchannel, at the output end $z = L$ indicates the fluctuation of its eigenvalue is similarly higher. This result, which we confirmed in our numerical simulations, is consistent with the theoretical prediction in Ref.~\cite{1990_Pichard}.
	
	The experimentally observed fluctuations of individual transmission eigenchannels are quantitatively reproduced by the numerical simulations of $t_{\rm buff\rightarrow end}$ and $t$ in Figs.~\ref{fluctuations}(a,b,d-f). The excellent agreement between experimental and numerical results confirms that eigenchannel fluctuations depend on their transmittance. The higher the transmittance, the lower the fluctuations. This means that high-transmission eigenchannels have a robust and consistent depth profile: irrespective of the disorder configuration of a system.
	
	\begin{figure}
		\centering
		\includegraphics[width=\linewidth]{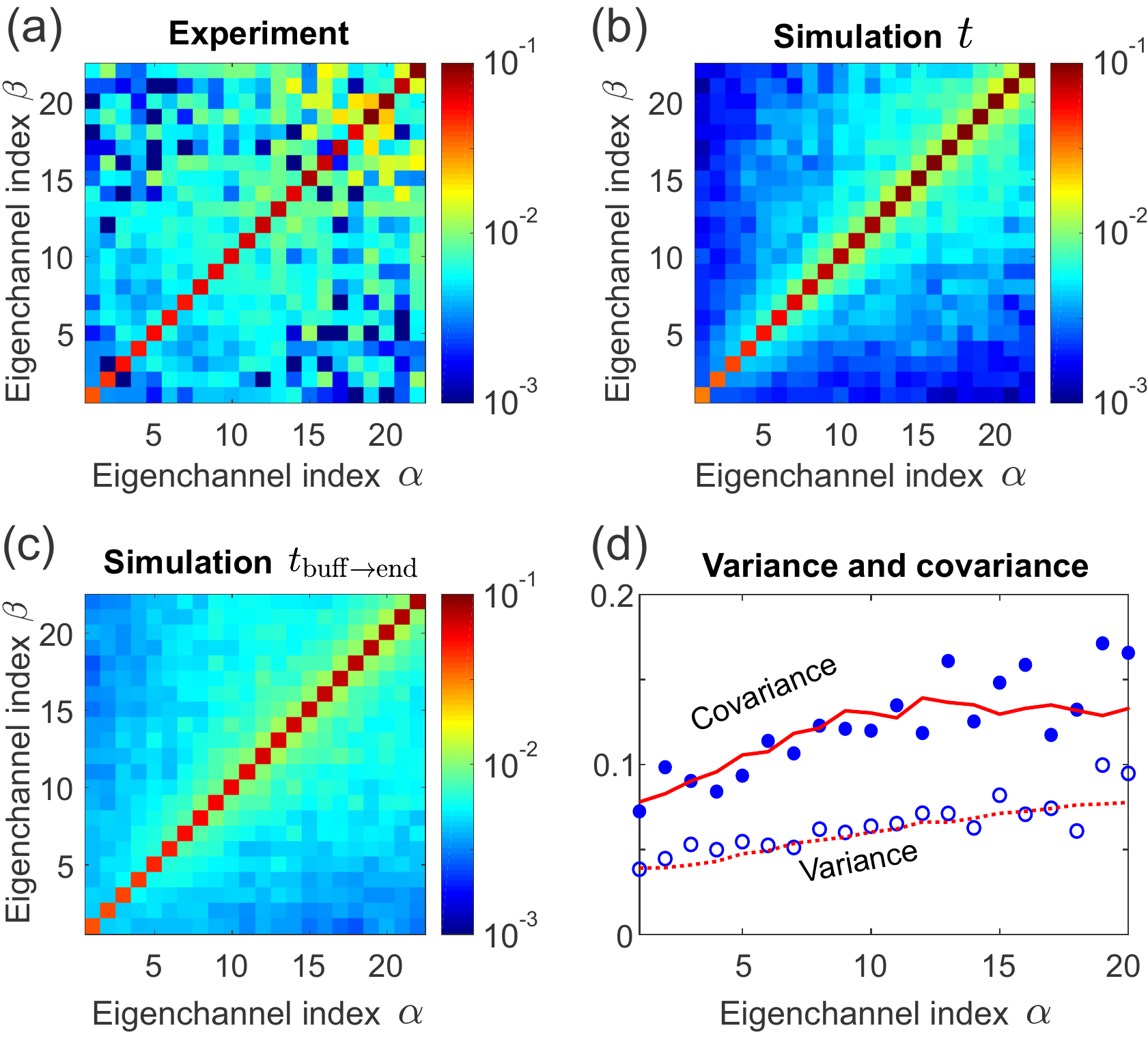}
		\caption{\label{correlations} 
			Inter-channel correlations. The covariance $\tilde{C}_{\alpha\beta}$ between any two pairs of eigenchannels, $\alpha$ and $\beta$, is calculated from experimental data (a) and numerical simulations (b,c). The cumulative covariance $\sum_{\beta\neq\alpha}\tilde{C}_{\alpha\beta}$ exceeds the variance $\tilde{C}_{\alpha \alpha}$ in (d). The blue symbols represent experimental data and red lines represent numerical simulations based on $t_{\rm buff\rightarrow end}$.
		} 
	\end{figure}
	
	Finally, we investigate the cross-correlations between different transmission eigenchannels. For any given disorder realization, eigenchannels are an orthogonal set of functions at the front and back surfaces of the medium. While eigenchannels differ from realization-to-realization, their orthogonality implies that the differences in their field profiles must be correlated from realization-to-realization. This does not mean, however, that the {\it intensity} fluctuations of their profiles {\it inside} the sample should be correlated. To study cross-correlations in the eigenchannels' intensity fluctuations across the sample, we introduce the covariance $\tilde{C}_{\alpha\beta}=\langle \delta I_\alpha(z) \delta I_\beta(z) \rangle_z$, where $\langle ...\rangle_z$ describes both ensemble averaging and depth averaging. For $\alpha=\beta$, $\tilde{C}_{\alpha\alpha}$ reduces to the variance $\tilde{C}_{\alpha}$ which describes the eigenchannel fluctuations.
	
	Fig.~\ref{correlations}(a-c) shows the experimental and numerical results of $\tilde{C}_{\alpha\beta}$ for all $\alpha$ and $\beta$. The non-vanishing off-diagonal elements of $\tilde{C}_{\alpha\beta}$ ($\alpha\neq\beta$) reveal coordinated changes in the eigenchannels' depth profiles. Between different pairings of eigenchannels, the correlations differ. The larger the difference in the transmittances of a pair, the weaker the correlation of their depth profile fluctuations. Furthermore, lower-transmission eigenchannels tend to correlate more with other low-transmission eigenchannels than higher-transmission eigenchannels do with other high-transmission eigenchannels. Quantitatively we can describe the correlation of a single eigenchannel to all others by the cumulative covariance $\sum_{\beta\neq\alpha}\tilde{C}_{\alpha\beta}$. As shown in Fig.~\ref{correlations}(d), the cumulative covariance increases with $\alpha$, indicating higher-transmission eigenchannels are more independent from other eigenchannels than lower-transmission eigenchannels. Moreover, the cumulative covariance exceeds the variance $\tilde{C}_{\alpha\alpha} = \tilde{C}_{\alpha}$ by a factor of 2. Hence, the total cross-correlation for a single eigenchannel is stronger than its own fluctuation. 
	
    To provide a plausible explanation for the observed phenomena, we resort to the modal description of transmission eigenchannels~\cite{shi2015dynamic}. A transmission eigenchannel can be decomposed by the quasinormal modes of the disordered system. Previous research~\cite{shi2015dynamic} has revealed that high-transmission eigenchannels are composed of only a few on-resonance modes, while low-transmission eigenchannels are composed of many off-resonance modes that destructively interfere. Since the destructive interference is sensitive to changes in the scattering configuration, the low-transmission eigenchannels exhibit strong fluctuations. Moreover, because individual low-transmission eigenchannels share many of the same off-resonant modes, their fluctuations are correlated. Since high-transmission eigenchannels are composed of a different set of modes than low-transmission eigenchannels, the correlations between high and low-transmission eigenchannels are weak.	
		
    Our findings regarding the second-order statistical properties of transmission eigenchannels are general and applicable to other types of waves such as microwaves, acoustic waves, and matter waves. In practical applications, the consistent and robust depth profiles of open channels guarantee that they can deliver energy deep into {\it any} diffusive system regardless of the disorder configuration. Such {\it reliable} energy delivery has major implications in applications ranging from multi-photon imaging to photothermal therapy, and shock wave treatment.
    Since our on-chip experimental platform allows for both direct measurement of the complex field inside a random structure and near-complete control over the incident field, we can investigate how to shape an incident wavefront to control the spatial distribution of light across the entire disordered sample. Furthermore, this setup can be used to experimentally study the spatial structure and statistics of  the time-delay eigenchannels of a diffusive system, as well as the time-gated transmission and reflection eigenchannels of a diffusive system.

	\begin{acknowledgments}
		
		This work is supported partly by the Office of Naval Research (ONR) under Grant No. N00014-20-1-2197, and by the National Science Foundation under Grant Nos. DMR-1905465, DMR-1905442 and OAC-1919789. 
		
	\end{acknowledgments}
	
	\bibliography{2020_Eigenchannel_correlations}

\end{document}


\preprint{APS/123-QED}

\title{Supplementary: Fluctuations and correlations of transmission eigenchannels in diffusive media}

\author{Nicholas Bender$^{1}$}
\thanks{These two authors contributed equally.}
	
\author{Alexey Yamilov$^{2},^{*}$}
\email{yamilov@mst.edu}

\author{Hasan Y\i lmaz$^{1}$}
	
\author{Hui Cao$^{1}$}
\email{hui.cao@yale.edu}

\affiliation{Department of Applied Physics, Yale University, New Haven, Connecticut 06520, USA \\
Physics Department, Missouri University of Science \& Technology, Rolla, Missouri 65409, USA}

\date{\today}
	
	\begin{abstract}

	\end{abstract}
	
	\maketitle

\section{Sample design and fabrication}

\begin{figure}[ht]
\centering
\includegraphics[width=\linewidth]{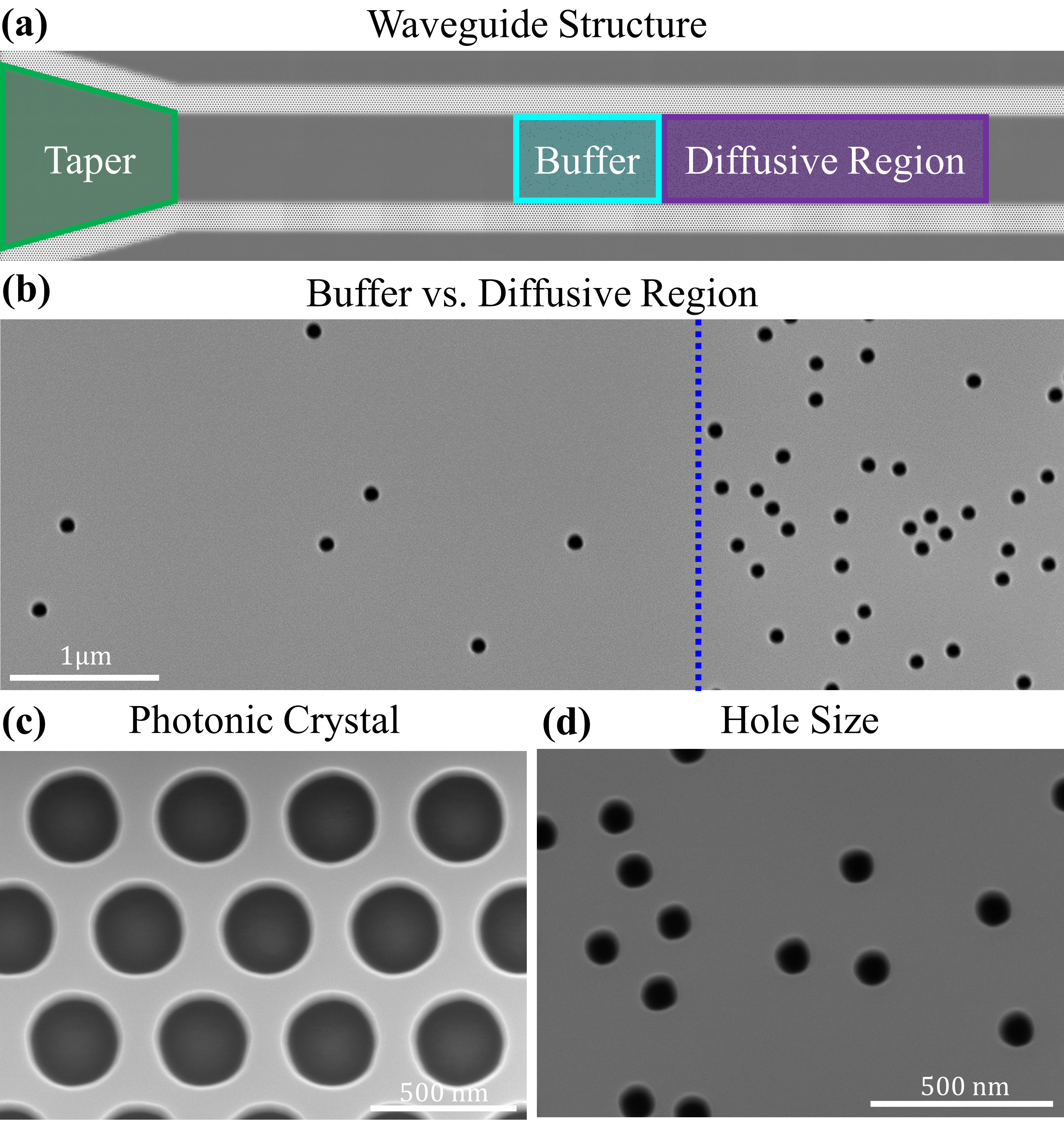}
\caption{\label{fig:SEM} 
Scanning electron microscope (SEM) images of a 2D diffusive waveguide. In (a) we show a composite SEM image which outlines the structures we etch into a silicon-on-insulator wafer when fabricating our structures. A SEM image of the interface between the buffer and diffusive regions is marked by the blue dashed line in (b). Close-up images of the photonic crystal sidewall and randomly-distributed holes are shown in (c) and (d).
} 
\end{figure}

In our previous work \cite{SarmaPRL16}, we optimized the SLM phase pattern to maximize or minimize the ratio of the light intensities, spatially integrated, in the back and front regions of a diffusive waveguide. This method resulted in simultaneous excitation of {\it multiple} high or low-transmission eigenchannels: but not a single eigenchannel. To measure the fluctuations of the individual eigenchannel profiles, we must excite {\it only one} eigenchannel at a time and measure its intensity profile. This requires experimentally measuring the field transmission matrix of the on-chip diffusive waveguide. It is important to point out, that even measuring the transmission matrix of light from the spatial light modulator (SLM) to the end of a diffusive waveguide is not sufficient for our purposes. This matrix not only includes information about the light transport through the disordered region in the waveguide, but also the optical transmission from the SLM to the edge of the wafer, the coupling of light into the lead waveguide, and the subsequent propagation through a tapered segment before reaching the disordered region.

In order to measure the transmission matrix just for the disordered region, we must access the incoming field right before this region. To this end, we introduce a weakly scattering “buffer region” into our samples, in front of the diffusive waveguide. By adjusting the air hole size and density in the buffer region, we are able to obtain just enough out-of-plane scattering to gain information about incoming waves while maintaining low loss and near complete control of the incoming wavefront.

Figure~\ref{fig:SEM} shows a schematic of our two-dimensional (2D) disordered waveguide structures. The major components are the tapered waveguide, the buffer region, and the diffusive region. The air holes (diameter = 100 nm), which induce light scattering in the buffer and diffusive regions, are randomly distributed with a minimum (edge-to-edge) distance of 50 nm. The diffusive region has 5250 holes, which results in an air filling fraction in the Si of $5.5 \% $. The number of air holes in the buffer region is 260, and the air filling fraction is $0.55 \% $. The sidewalls of the waveguide consist of a trigonal lattice of air holes (radius = 155 nm, lattice constant = 440 nm). They provide a 2D complete bandgap for TE polarized light (used in the experiment) within the wavelength range of 1120 nm to 1580 nm~\cite{2014_Yamilov_Dofz_experiment}.

The probe light is injected from the side/edge of the wafer into a ridge waveguide (width = 300 \textmu m, length = 15 mm). It then enters a tapered waveguide (tapering angle = $15^{\circ}$). The tapered waveguide width decreases gradually from 300 \textmu m to 15 \textmu m. The tapering results in waveguide mode coupling and conversion~\cite{SarmaPRL16}. To avoid light leakage, the tapered waveguide has photonic crystal sidewalls.

\section{Optical Setup}

\begin{figure}[ht]
\centering
\includegraphics[width=\linewidth]{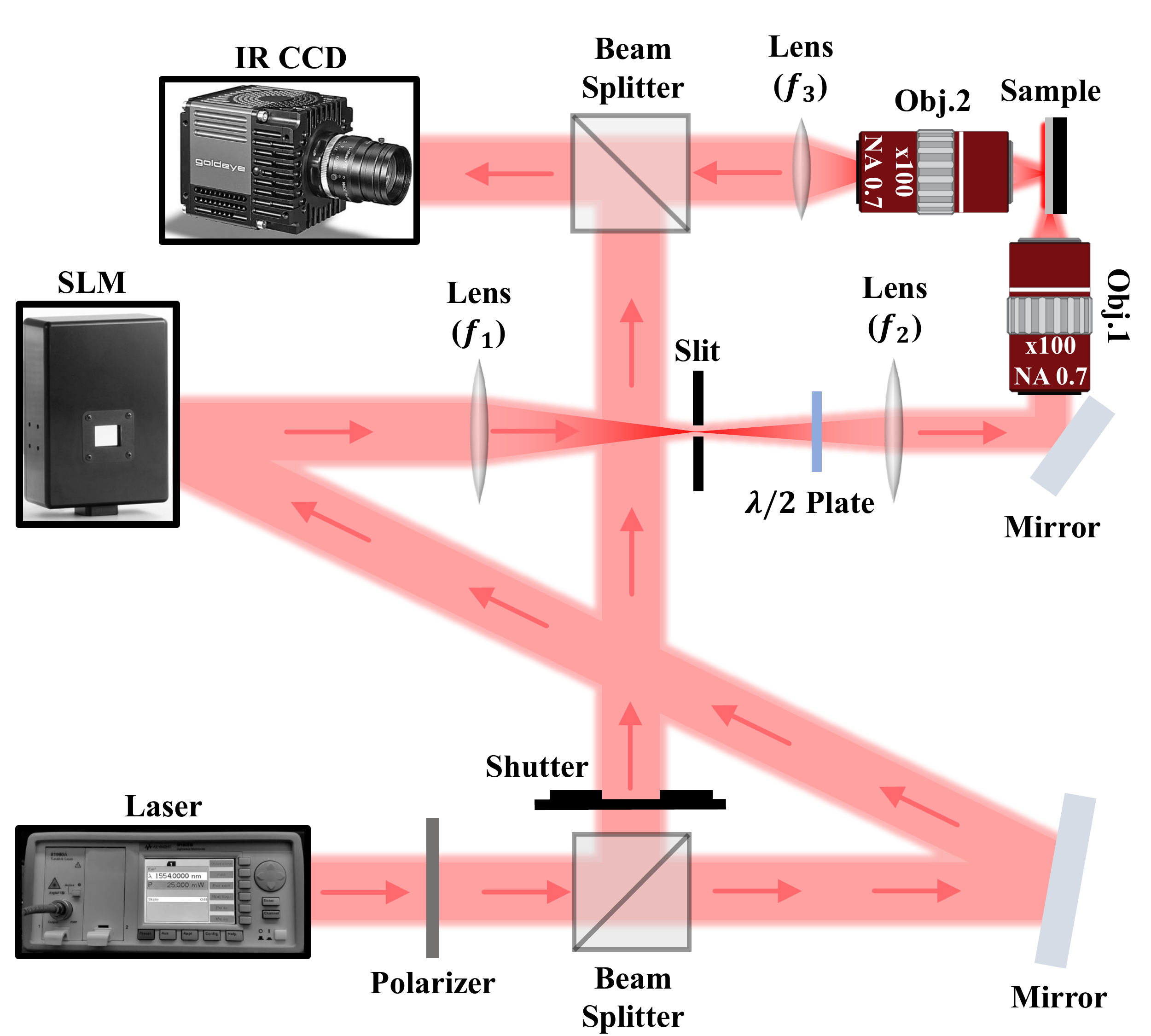}
\caption{\label{fig:setup} 
A depiction of our experimental setup. Monochromatic light from our laser is linearly polarized and split into two beams. One beam illuminates the phase modulating surface of a spatial light modulator (SLM), while the other is used as a reference beam. The SLM is used to control the input wavefront in our diffusive waveguide structures. A beam splitter merges the light collected from the top of our sample with the reference beam on an IR CCD. The focal length of the three lenses used in this setup are: $f_{1}=400$ mm, $f_{2}=75$ mm, and $f_{3}=100$ mm.
} 
\end{figure}

Fig.~\ref{fig:setup} is a detailed schematic of the experimental setup. Continuous-wave (CW) output from a tunable laser (Keysight 81960A) -operating around $1554$ nm- is linearly polarized and split into two beams. One beam illuminates the phase modulating surface of a phase-only SLM (Hamamatsu LCoS X10468), while the other is used as a reference beam. A one-dimensional (1D) phase-modulation pattern is displayed on the SLM, consisting of 128 macropixels. Each macropixel consists of $4\times800$ regular pixels on the SLM. Using two lenses with focal lengths of $f_{1}=400$ mm and $f_{2}=75$ mm, we image the field on the SLM plane onto the back focal plane of a long-working-distance objective ({\bf Obj. 1}) (Mitutoyo M Plan APO NIR HR100$\times$, Numerical Aperture = 0.7). To prevent the unmodulated light from entering the objective lens, we display a binary diffraction grating within each macropixel to shift the modulated light away from the unmodulated light in the focal plane of the $f_{1}$ lens. Using a slit in the same focal plane, we block everything except the phase-modulated light in the first diffraction order. Before the the $f_{2}$ lens, we insert a half-wave $(\lambda/2)$ plate to flip the polarization of light so that it is TE polarized relative to our waveguide sample. The side of our SOI wafer is placed at the front focal plane of {\bf Obj. 1} and illuminated with the Fourier transform of the phase-modulation pattern displayed on the SLM. From the top of the wafer, a second long-working-distance objective ({\bf Obj. 2}) (Mitutoyo M Plan APO NIR HR100$\times$) collects light scattered out-of-plane from the waveguide. We use a third lens with a focal length of $f_{3}=100$ mm together with {\bf Obj. 2} to magnify the sample image by $\times 50$. With a second beam splitter, we combine the light collected from the sample and the reference beam. Their interference patterns are recorded with an IR CCD camera (Allied Vision Goldeye G-032 Cool).

\section{Interferometric Measurement}

\begin{figure}
	\begin{center}
		\includegraphics[width=\linewidth]{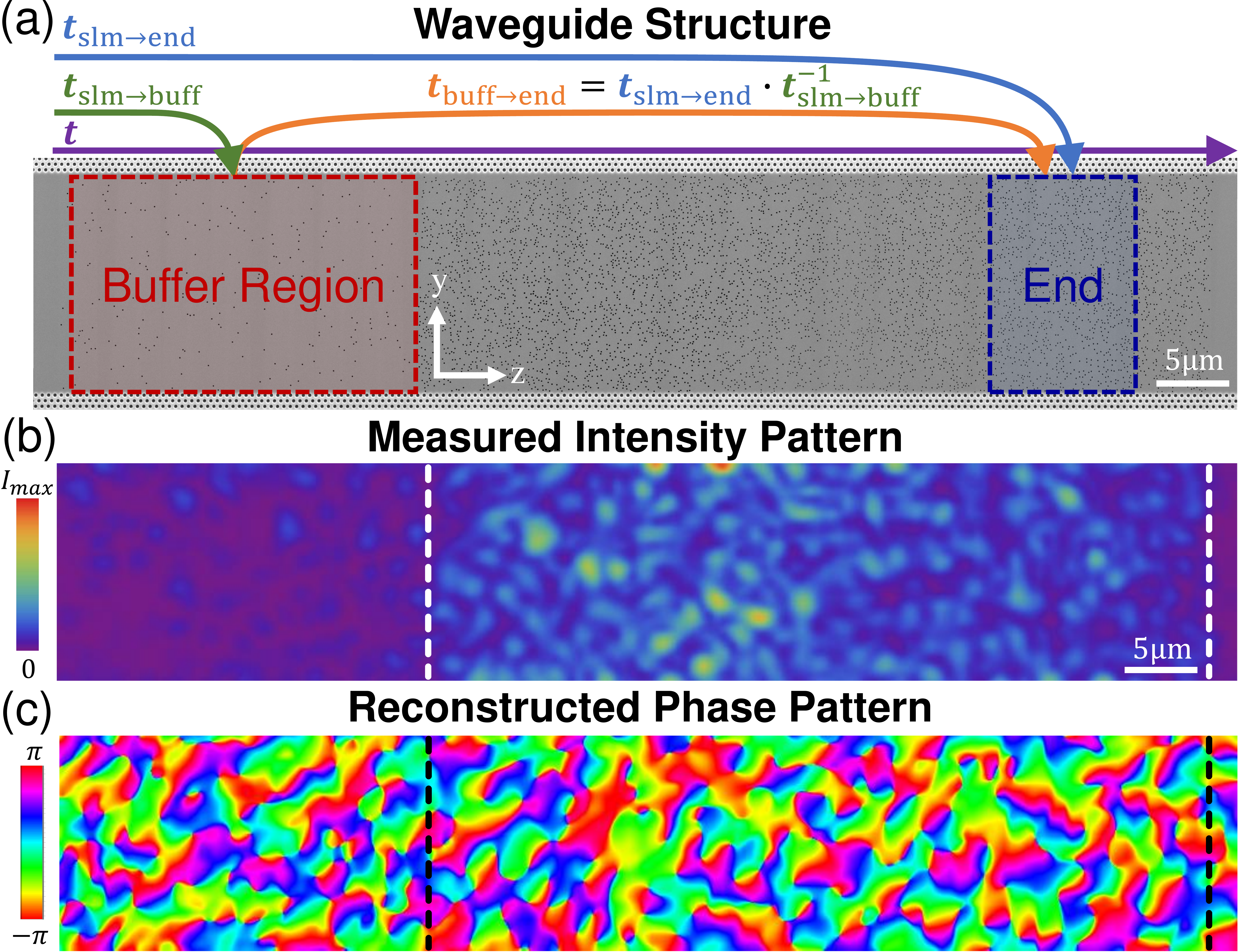}
		\caption{\label{fig:fullfield} 
			Waveguide structure and full-field measurement. A composite SEM image of a diffusive waveguide is shown in (a). In (b) the 2D intensity pattern of a measured high-transmission eigenchannel is shown. Using our interferometric setup, we can reconstruct the phase of the light field  inside the diffusive waveguide in (c). In (b-c) the edges of the diffusive region are marked by the vertical dashed lines.
		} 
	\end{center}
\end{figure}

With the interferometric setup described in the last section, we can measure the field distribution of light scattered out-of-plane from within the diffusive waveguide: for any phase-modulation pattern displayed on the SLM. To do this, we first measure the 2D intensity distribution of the scattered light by blocking the reference beam with a shutter [Fig.\ref{fig:fullfield}(b)]. Then using the reference beam in our setup, we retrieve the phase profile of the scattered light with a four-phase measurement (wherein the global phase of the pattern on the SLM is modulated four times in increments of $\pi/2$ rad)~\cite{CCCL}. Fig.~\ref{fig:fullfield}(c) shows the spatial distribution of the recovered phase pattern of the light field across the diffusive waveguide.

By measuring the complex field throughout the waveguide for an orthogonal set of phase patterns displayed on the SLM, we can construct two matrices $t_{\rm slm\rightarrow buff}$ and $t_{\rm slm\rightarrow end}$, which map the field from the SLM surface to the buffer and to the far end of the disordered waveguide, respectively. To construct the matrix relating the field in the buffer region to the field near the end of the diffusive waveguide, $t_{\rm buff\rightarrow end}$, we define the field-mapping matrix between the two regions $t_{\rm buff\rightarrow end}\equiv t_{\rm slm\rightarrow end} \, t_{\rm slm\rightarrow buff}^{-1}$. To calculate the inverse of $t_{\rm slm\rightarrow buff}$, we use Moore-Penrose matrix inversion. In this operation we only take the inverse of the $55$ highest singular values of $t_{\rm slm\rightarrow buff}$, and set the inverse of the remaining singular values to zero. This restriction is imposed because our diffusive waveguide only has 55 transmission eigenchannels.

\section{\label{sec:fitting} Determination of transport parameters}
Diffusive wave propagation in a scattering medium with loss is determined by two parameters: the transport mean free path $\ell_t$ and the diffusive dissipation length $\xi_a$. In a 2D system, the latter can be expressed as $\xi_a=\sqrt{\ell_t \ell_a/2}$, where $\ell_a$ is the ballistic dissipation length. 

To determine $\xi_a$ and $\ell_t$ in the diffusive region of the 2D waveguide, we first measure the cross-section-averaged intensity  $I(z)$ as a function of depth $z$ for multiple random input wavefronts. We then ensemble average the data, $\langle I(z)\rangle$, and fit the theoretically-predicted depth profile -based on the diffusive equation- to it. The theoretical $\langle I(z)\rangle$ is found by convolving the incident ballistic intensity $I_0\exp[-z/\ell_s]$ ($\ell_s$ represents the scattering mean free path, and $\ell_s\approx \ell_t$ in our case), which acts as the source, and the Green's function of the diffusion equation~\cite{1993_Lisyansky_diffusint}:
\begin{equation}
   G(z,z^\prime)=\left\{ 
    \begin{array}{c}
    P(z)P(L-z^\prime),\ z<z^\prime \\
    P(z^\prime)P(L-z),\ z>z^\prime \\
    \end{array}
   \right.
\end{equation}
where $P(z)=\sinh(z/\xi_a)+z_0/\xi_a\cosh(z/\xi_a)$, and $z_0=(\pi/4)\times\ell_t$ is the so-called extrapolation length.

\begin{figure}[ht]
\centering
\includegraphics[width=2.7in]{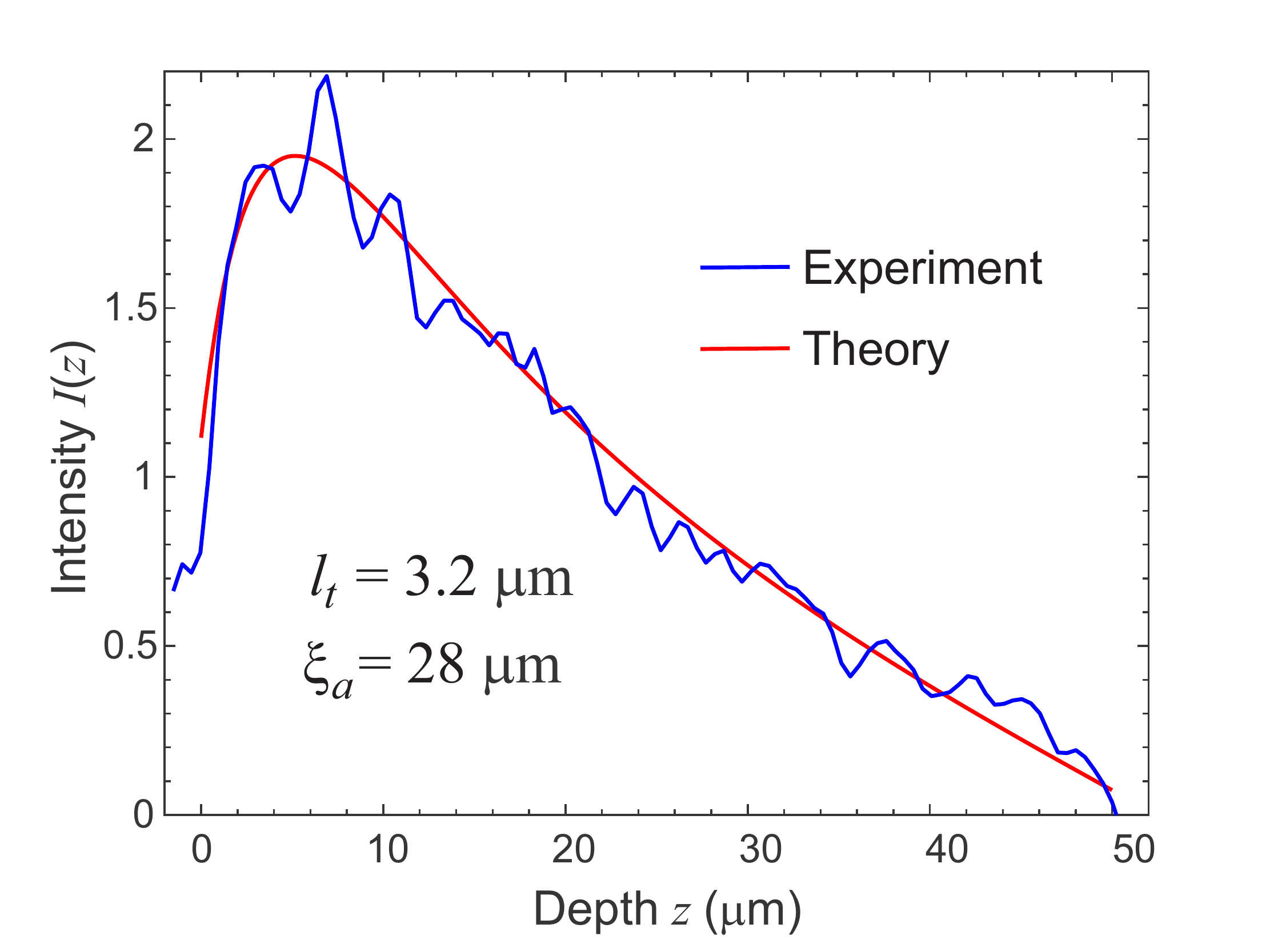}
\caption{\label{fig:fit} 
Determining the transport mean free path $\ell_t$ and the diffusive dissipation length $\xi_a$ of the diffusive waveguides by fitting the experimentally-measured average depth profile for random incident wavefronts, $\langle I(z)\rangle$, (blue) to theoretical predictions from the diffusion equation (red).
} 
\end{figure}

We compute the difference between the experimental and theoretical $\langle I(z) \rangle$ for different values of $\xi_a$ and $\ell_t$, and identify the minimum  difference at $\xi_a=28$ \textmu m and $\ell_t=3.2$ \textmu m. Fig.~\ref{fig:fit} shows an excellent agreement between the measured $\langle I(z) \rangle$ and the theoretical prediction. 

In the buffer region, the air hole density is 10 times lower than in the diffusive region. Thus, the transport mean free path is 10 times longer, $\ell_t^{\rm buff}=32$ \textmu m. The loss, caused by out-of-plane scattering from the air holes, is also 10 times weaker, thus the ballistic dissipation length $\ell_a$ is 10 times longer. This leads to a tenfold increase in the diffusive dissipation length: $\xi_a^{\rm buff}=280$ \textmu m.

\section{Eigenchannel Profile Measurement}
In total, we measure the transmission eigenchannel intensity profiles of 13 independent realizations. We obtain these measurements from two samples with different random hole configurations. To generate independent system realizations from the same random hole configuration, we vary the wavelength of the input light beyond the spectral correlation width of the diffusive region: 0.4 nm. Over a wavelength span of 3 nm, we vary the input wavelength of our laser in increments of 0.5 nm. We choose the specific wavelength range of the measurement -for each random hole configuration- such that the effective dissipation in the diffusive region is minimal over the wavelength range and homogeneous. While our waveguide structure has a width of 15 \textmu m, we only use the central 10 \textmu m region of the waveguide’s out-of-plane-scattered light when performing our measurements to avoid artifacts from light scattered out-of-plane from the photonic crystal boundaries.

\section{\label{sec:simulations} Numerical simulations}
In our numerical simulations, we use the recursive Green's function method in the Kwant simulation package~\cite{2014_Groth_Kwant}. We simulate a two-dimensional (2D) rectangular waveguide geometry, which is defined using a tight-binding model for scalar waves on a square grid. At the waveguide boundaries, which are reflective, the grid is terminated. The leads are attached to the open ends of the waveguide, allowing for computation of the complete scattering $S$ matrix of the system and the wave field throughout the bulk of the system: under an excitation by an arbitrary combination of field amplitudes for the propagating modes. The width $W$ of the simulated system is selected so that the number of waveguide modes $N$  matches the number found in the experiment. Once $W$ is chosen, the length of the disordered waveguide is determined by the ratio $L/W$ of the waveguides used in the experiment. Due to the low filling fraction of the air holes in the experimental waveguides, both in the buffer region and in the main disordered region, we assume that the number of propagating modes is equal to $N$. 

Scattering is introduced by a randomly (box distribution) fluctuating real-valued on-site `energy' in the tight-binding model, see Refs.~\cite{SarmaPRL16, 2016_Yamilov_Eigenvalues}. The addition of a positive imaginary constant to the same term simulates the effect of absorption. In our previous works, we confirmed that the process of vertical leakage due to the holes in our disordered waveguides can be modeled via absorption in a 2D system~\cite{2014_Yamilov_Dofz_experiment, 2014_Sarma_Cofz, SarmaPRL16}. The actual material absorption in our experimental system is negligible. By a proper choice of these parameters, we can match the experimental values for the transport mean free path $\ell_t$ and the diffusive dissipation length $\xi_a$. 

To model the weakly scattering `buffer' region, we reduce the scattering (the amplitude of the on-site fluctuation) so that transport mean free path is reduced by a factor of $10$. The latter corresponds to a 10 times reduction in the areal density of the air holes in the buffer region. Furthermore, because the out-of-plane scattering loss is reduced 10 times, the diffusive dissipation length is also reduced by the same factor. 

The buffer region is incorporated into the experimental waveguides to measure $t_{\rm buff\rightarrow end}$ of the diffusive waveguide, which is not a direct measurement of the field transmission matrix $t$. We numerically simulate the eigenchanels of both matrices to confirm their depth profiles are equivalent. The matrix $t$ is obtained from the incident and transmitted fields in the left and right leads without the buffer. To compute $t_{\rm buff\rightarrow end}$, we compute the auxiliary matrices $t_{\rm in\rightarrow buff}$ and $t_{\rm in\rightarrow end}$. The former matrix relates the incident fields in the left lead to the fields at $2\times N$ randomly selected points within a 10 \textmu m $\times$ 20 \textmu m region centered in the buffer region (of an area 15 \textmu m $\times$ 25 \textmu m). The chosen points are at least 2.5 \textmu m separate from each other or any boundary/interface. The second auxiliary matrix  $t_{\rm in\rightarrow end}$ relates the impinging fields in the left lead to the fields at $2\times N$ randomly selected points within a 10 \textmu m $\times$ 10 \textmu m region at the end of the diffusive waveguide. Again all points are at least 2.5 \textmu m (which is on the order of $\ell_t$) spaced. In the last step, we compute $t_{\rm buff\rightarrow end}= t_{\rm in\rightarrow end}t_{\rm in\rightarrow buff}^{-1}$, where $t_{\rm in\rightarrow buff}^{-1}$ is calculated with the Moore-Penrose pseudo-inverse.

To calculate the spatial structure of the transmission eigenchannels, we perform a singular value decomposition on the $t$ matrix, and use the right singular vectors as input fields in the left lead to excite individual eigenchannels. For the matrix $t_{\rm buff\rightarrow end}$, its right singular vectors are transformed to the incident fields in the left lead by multiplying $t_{\rm in\rightarrow buff}^{-1}$. To further mimic the phase-only modulation of the SLM in the experiment, we only keep the phases of the incident fields, and set the field magnitudes equal. We calculate all eigenchannels for $t$ and $t_{\rm buff\rightarrow end}$ for an ensemble of $1000$ disorder configurations of the waveguides. The numerical results are presented in Figs.~2-4 in the main text.

To compare the variance $ \tilde{C}_{\alpha}$ and covariance $\tilde{C}_{\alpha \beta}$ numerically calculated from $t_{\rm buff\rightarrow end}$ to the experimentally-measured ones, we need to account for some experimental limitations and imperfections. On one hand, the finite spatial resolution of our detection optics effectively enlarges the speckle grain size of the field measured inside the diffusive waveguide. This reduction in the number of speckle grains increases the fluctuations of the cross-section-averaged intensity. On the other hand, the combined effects of sample drift during measurements and the presence of two linear polarizations in the light scattered out-of-plane from our sample; decrease the fluctuations of the cross-section-averaged intensity. For random incident wavefronts, the spatially-averaged intensity variance of our experimental measurements is $ {\rm var}[I (z)] = 0.59$, compared to $ {\rm var}[I (z)] = 0.64$ from the numerical simulations of $t_{\rm buff\rightarrow end}$. For all eigenchannels, we re-scale the numerical ${\rm var}[I_\alpha (z)]$ and $\tilde{C}_{\alpha\beta}$ by the multiplicative factor $0.59/0.64$, in order to compare them to the experimental values. While we applied the re-scaling factor to the fluctuations and correlations calculated from numerical simulations of $t_{\rm buff\rightarrow end}$, we did not apply it to the results from simulations of $t$.

\begin{figure}[h]
\centering
\includegraphics[width=2.7in]{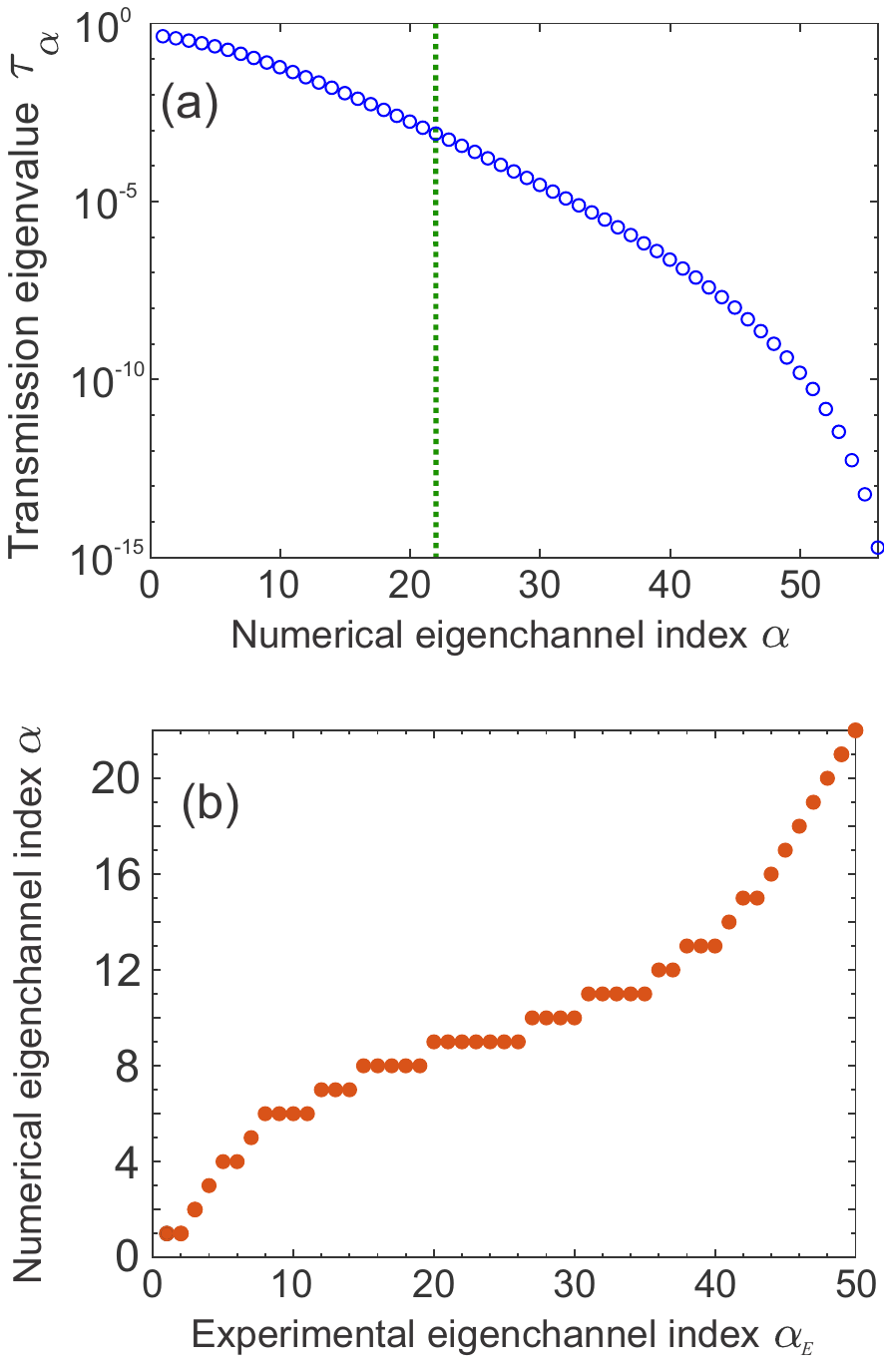}
\caption{\label{fig:mapping} 
Calculated transmission eigenvalues, as a function of eigenchannel index $\alpha$, are shown in (a). In (b), we show the mapping between the experimentally-measured eigenchannel profiles with index $\alpha_E$ and the first $22$ (in the order of decreasing transmittance) eigenchannels with index $\alpha$ found in the numerical simulations based on the $t_{\rm buff\rightarrow end}$ matrix.}
\end{figure}

\section{\label{sec:eigenchannel_id} Identification of experimental eigenchannels}

In this section, we analyze the normalized eigenchannel profiles measured in the experiment $\langle I_{\alpha_E}(z)\rangle$ and the numerical simulations $\langle I_{\alpha}(z)\rangle$. For each experimental eigenchannel with an index of $\alpha_E\in [1...55]$, we identify the corresponding numerical eigenchannel with an index that minimizes the difference $\int_0^L(\langle I_{\alpha_E}(z)\rangle-\langle I_{\alpha}(z)\rangle)^2 \, dz$. We do \textit{not} use any eigenchannel-specific adjustments/fits in this identification. This process gives the mapping of $\alpha_E$ to $\alpha$, shown in Fig.~\ref{fig:mapping}. A few experimental eigenchannels are redundant, particularly in the range $\alpha\in[6...15]$, and no eigenchannels with $\alpha>22$ are observed experimentally. We attribute this to the finite signal-to-noise ratio in the experimental data. The eigenchannels with $\alpha>22$ have a transmittance less than $\sim 0.25\%$, thus they are overwhelmed by the experimental noise. 

We use the redundancy of the experimental eigenchannels in Fig.~\ref{fig:mapping} to enlarge the statistical ensemble. In other words, statistical averages $\langle...\rangle$ for the $\alpha$-th eigenchannel that corresponds to multiple $\alpha_E$'s include both disorder configuration and $\{\alpha_E\}$ averaging.

\section{\label{sec:var} Eigenchannel variance}

\begin{figure}[htb]
	\centering
	\includegraphics[width=2.7in]{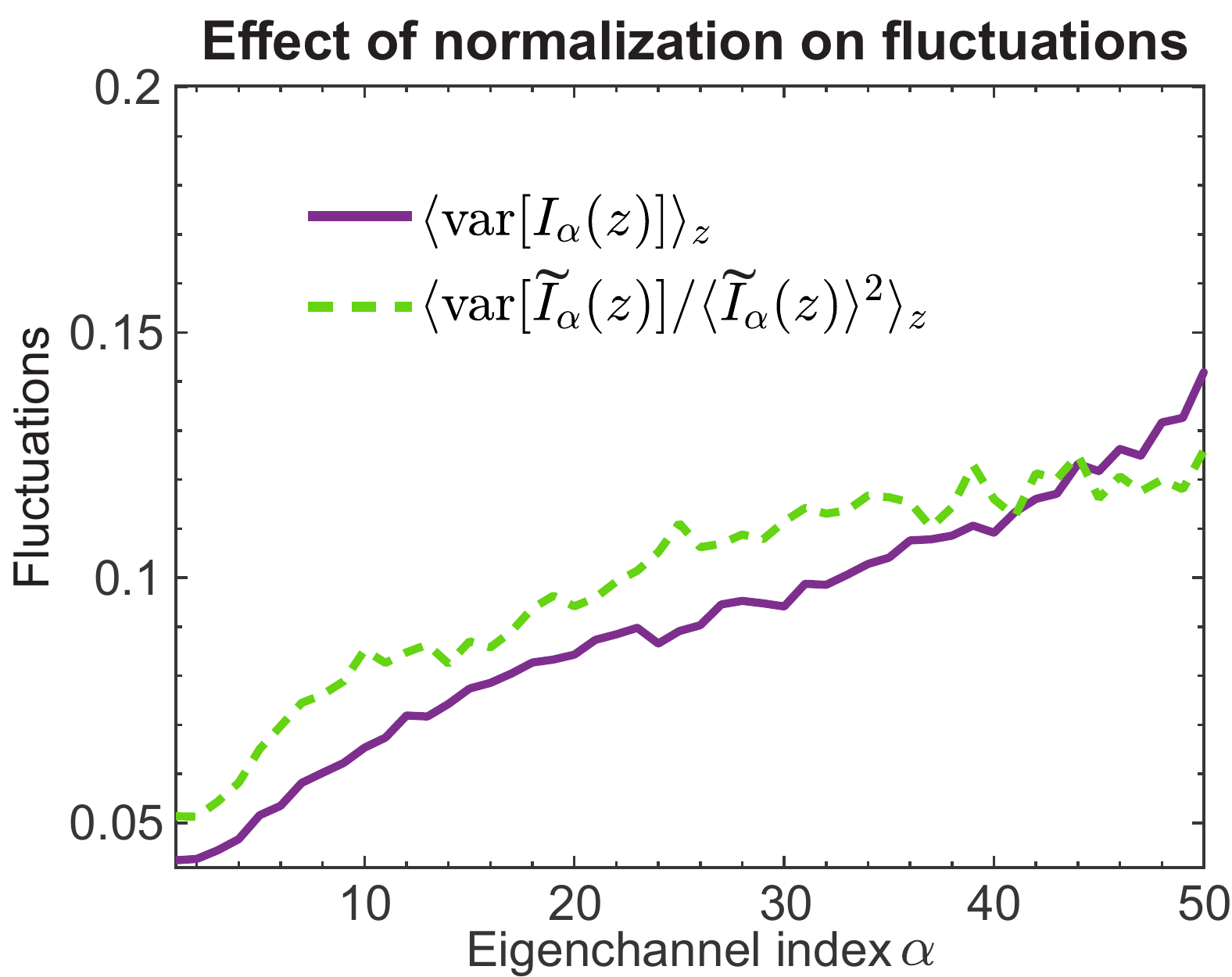}
	\caption{\label{variance} We compare the variance of eigenchannel profiles calculated with the normalized intensity (purple solid line) to the intensity variance normalized by the mean intensity $\langle {\rm var}[\tilde{I}_\alpha (z)] / \langle \tilde{I}_{\alpha}(z) \rangle^2 \rangle_z$ squared (green dashed line). They show similar growth with the eigenchannel index $\alpha$.}
\end{figure}

In the main text, we present the realization-to-realization fluctuations of the egenchannels’ normalized intensity profiles. For each eigenchannel, the measured intensity profile $\tilde{I}(z)$ is normalized to $I(z)=\tilde{I}(z)/ [(1/L)\int_0^L \tilde{I}(z')dz']$. Using a different normalization procedure, we check the effect of our normalization on the eigenchannel fluctuations using numerical simulations of $t$. For an eigenchannel $\alpha$, the variance $ {\rm var}[\tilde{I}_\alpha (z)] = \langle \delta \tilde{I}_{\alpha}^2(z) \rangle $ of the unnormalized intensity fluctuation $\delta \tilde{I}_{\alpha}(z) = \tilde{I}_{\alpha}(z) - \langle \tilde{I}_{\alpha}(z) \rangle$ can be normalized by dividing the square of the mean intensity $\langle \tilde{I}_{\alpha}(z) \rangle^2$ at the same depth $z$. Then this ratio ${\rm var}[\tilde{I}_\alpha (z)] / \langle \tilde{I}_{\alpha}(z) \rangle^2$ can be averaged over all $z$. In Fig.~\ref{variance}, we compare this quantity to the variance of the normalized intensity profile, $\tilde{C}_{\alpha}$, calculated in the main text. Both exhibit an increase with the eigenchannel index $\alpha$. Their similar trend confirms that the stronger fluctuations for lower-transmission eigenchannels are due to the intrinsic properties of the transmission eigenchannels.    

\bibliography{2020_Eigenchannel_correlations}